\documentstyle[pra,aps,amssymb]{revtex}

\newcommand{\nc}{\newcommand}
\nc{\vare}{\varepsilon} \nc{\oi}{{0i}}
\nc{\psib}{\overline{\psi}} \nc{\psid}{{\psi^{\dagger}}}
\nc{\bb}{\begin{equation}} \nc{\ee}{\end{equation}}
\nc{\erm}{{\rm e}} \nc{\munu}{{\mu\nu}}
\nc{\dis}{\displaystyle} \nc{\um}{{1\over 2}}
\nc{\R}{{\rm I\!R}} \nc{\C}{{\rm I\!\!\!C}}
\nc{\vecna}{\mbox{\boldmath $\nabla$}}
\nc{\pa}{\partial} \nc{\ug}{\; = \;} \nc{\vs}{\vspace*}
\nc{\Hc}{{\cal H}}  \nc{\Lc}{{\cal L}}  \nc{\Lcn}{{\cal L}^{(n)}}
\nc{\Lcuno}{{\cal L}^{(1)}}  \nc{\Lczero}{{\cal L}^{(0)}}
\nc{\Vbf}{\mbox{\boldmath $V$}} \nc{\Fbf}{\mbox{\boldmath $F$}}
\nc{\Wbf}{\mbox{\boldmath $W$}} \nc{\lbf}{\mbox{\boldmath $l$}}
\nc{\xbf}{\mbox{\boldmath $x$}} \nc{\ubf}{\mbox{\boldmath $u$}}
\nc{\vbf}{\mbox{\boldmath $v$}} \nc{\wbf}{\mbox{\boldmath $w$}}
\nc{\jbf}{\mbox{\boldmath $j$}}
\nc{\sigbf}{\mbox{\boldmath $\sigma$}}
\nc{\abf}{\mbox{\boldmath $a$}} \nc{\bbf}{\mbox{\boldmath $b$}}
\nc{\sbf}{\mbox{\boldmath $s$}} \nc{\rbf}{\mbox{\boldmath $r$}}
\nc{\kbf}{\mbox{\boldmath $k$}} \nc{\Lbf}{\mbox{\boldmath $L$}}
\nc{\imp}{\mbox{\boldmath $p$}} \nc{\albf}{\mbox{\boldmath $\alpha$}}
\nc{\qbf}{\mbox{\boldmath$q$}} \nc{\pibf}{\mbox{\boldmath$\pi$}}
\nc{\dddov}{{\stackrel{\ldots}{v}}} \nc{\dox}{\dot{x}} \nc{\ddox}{\ddot{x}}
\nc{\dddox}{{\stackrel{\ldots}{x}}} \nc{\dopi}{\dot{\pi}} \nc{\dop}{\dot{p}}
\nc{\dov}{\dot{v}} \nc{\ddov}{\ddot{v}} \nc{\doa}{\dot{a}} \nc{\ddoa}{\ddot{a}}
\nc{\dddoa}{{\stackrel{\ldots}{a}}} \nc{\ddddoa}{{\stackrel{....}{a}}}
\nc{\ddddov}{{\stackrel{....}{v}}}
\nc{\doS}{\dot{S}} \nc{\ddoS}{\ddot{S}} \nc{\doJ}{\dot{J}}
\nc{\doL}{\dot{L}} \nc{\ddoL}{\ddot{L}}
\nc{\doabf}{\mbox{\boldmath ${{\stackrel{.}{a}}}$}}
\nc{\ddoabf}{\mbox{\boldmath ${{\stackrel{..}{a}}}$}}
\nc{\doabfo}{\widehat{\doabf}}
\nc{\doao}{\widehat{\doa}}
\nc{\dopo}{\widehat{\dop}}
\nc{\doimp}{\dot{\imp}} \nc{\doq}{\dot{q}}
\nc{\ddoq}{\ddot{q}} \nc{\ddopi}{\ddot{\pi}}
\nc{\Wt}{{\widetilde{W}}} \nc{\doWt}{\dot{\Wt}}
\nc{\Omn}{\Omega^\munu} \nc{\po}{\widehat{p}}
\nc{\ga}{\gamma} \nc{\al}{\alpha}
\nc{\gabf}{\mbox{\boldmath $\gamma$}}
\nc{\rd}{{\rm d}} \nc{\dtau}{{\rd\tau}} \nc{\dt}{{\rd t}}
\nc{\cmf}{_{\star}} \nc{\para}{^{\parallel}} \nc{\orto}{^{\perp}}
\nc{\vi}{{v^{(\rm i)}}} \nc{\aii}{{a^{(\rm 2i)}}}
\nc{\M}{{\rm I\!\!M}} \nc{\ain}{{a^{(\rm 2n)}}}
\nc{\Ho}{\widehat{H}} \nc{\CoMF}{_{\rm {\footnotesize CMF}}}
\nc{\gao}{\gamma^0} \nc{\impo}{\widehat{\imp}}
\nc{\aoo}{\widehat{a}} \nc{\vo}{\widehat{v}}
\nc{\So}{\widehat{S}} \nc{\Go}{\widehat{G}}
\nc{\vbfo}{\widehat{\vbf}} \nc{\abfo}{\widehat{\abf}}
\nc{\Sigbf}{\mbox{\boldmath $\Sigma$}}
\nc{\Orm}{{\rm O}\!\!\!\!\!{\rm O}}
\nc{\ddoR}{\ddot{R}} \nc{\ddddoR}{{\stackrel{....}{R}}}
\nc{\Hco}{\widehat{H}}
\nc{\impl}{{ \ \ \ \Rightarrow \ \ \ }}
\nc{\Impl}{{ \ \ \ \Longrightarrow \ \ \ }}
\nc{\rdt}{\rd t} \nc{\nnmuno}{{NNM$^{(1)}$}\,}
\nc{\Sc}{{\cal S}} \nc{\un}{1\!\!1}
\nc{\lqq}{\left[} \nc{\rqq}{\right]}
\nc{\lt}{\left(} \nc{\rt}{\right)} \nc{\ik}{{ik}}
\nc{\eee}{\end{document}}

\begin{document}

\title{\rightline{\small{\em Published in} Int.J.Mod.Phys.{\bf A20}, 2027 (2005)}\vspace*{0.7cm}Non-relativistic 
classical mechanics for spinning particles\footnote{Work partially 
supported by I.N.F.N. and M.I.U.R}}
\author{\large Giovanni Salesi}
\address{Universit\`a Statale di Bergamo, Facolt\`a di Ingegneria,
Italy; {\em and}\\
Istituto Nazionale di Fisica Nucleare--Sezione di Milano,
Italy\footnote{e-mail: {\em salesi@unibg.it}}}
\maketitle
\vs{0.5 cm}
\begin{abstract}
\noindent We study the classical dynamics of non-relativistic particles
endowed with spin. Non-vanishing {\em Zitterbewegung} terms appear in the
equation of motion also in the small momentum limit.
We derive a generalized work-energy theorem which suggests classical
interpretations for tunnel effect and quantum potential.

\

\end{abstract}

\

\section{Spin and Zitterbewegung in Non-Newtonian Mechanics}

\noindent The basic kinematical feature of the motion of spinning
particles is the so-called {\em Zitterbewegung} [1--7], i.e.,
the high frequency jitter-motion first described by
Schr\"odinger \cite{Schroedinger} in the Thirties.
Because of Zitterbewegung
velocity and momentum of a spinning particle
are non-proportional but independent quantities
$$
\vbf \not\,\parallel \imp\,.
$$
Actually in Dirac theory the velocity and momentum operators are not
proportional
$$
\vbfo=\albf c \ \ \not\,\parallel \ \ \impo=-i\hbar\vecna\,.
$$
Furthermore $\vbfo$, differently from $\impo$, does not commute with
the Dirac Hamiltonian \mbox{$\Hco=\albf c\cdot\impo+\beta mc^2$} so that, while
$\impo$ is a constant quantity, $\vbfo$ is not.
Therefore {\em the dynamical structure of Quantum Mechanics is intrinsically
non-Newtonian}
$$
\vbf\neq\frac{\imp}{m} \qquad\qquad
\abf\neq\frac{1}{m}\frac{\rd\imp}{\rdt}=\frac{\Fbf}{m}\,.
$$
As a consequence Newton's Law and (in the absence of external
forces) Galileo's Principle of Inertia do not hold anymore, and the free
motion is not in general uniform rectilinear.
Zitterbewegung is well depicted through the celebrate Gordon
decomposition \cite{Gordon}
of the Dirac probability current:
$$
j^\mu=\psib\ga^\mu\psi=\frac{1}{2m}\,\lqq\psib(\po^\mu\psi) -
(\po^\mu\psib)\psi\rqq + \frac{1}{m}\pa_\nu\,\lt\psib\So^\munu\psi\rt
$$
[$\psib\equiv\psid\gao$, $\po^\mu \equiv i\hbar\pa^\mu$, and $\So^{\mu\nu}\equiv
i\hbar(\ga^\mu\ga^\nu - \ga^\nu\ga^\mu)/4$ represents the spin tensor operator].
The first term in the r.h.s.\,\,is associated with the translational motion of
the CM; whilst, the non-Newtonian term in the r.h.s.\,\,is related to the
existence of the spin, and describes the Zitterbewegung rotational motion.

Analogous Gordon-like decompositions of the conserved currents can be
written for spin-1 bosons and for \mbox{spin-${3\over 2}$} fermions (in
the Proca and
Rarita-Schwinger theories, respectively), as well as for NR particles,
in Pauli's and Schr\"odinger's theories\footnote{In fact, following
Landau \cite{Landau-1}, we can write a NR Gordon-like decomposition of
the conserved Pauli current
$$
\jbf = \frac{i\hbar}{2m}\lqq\lt\vecna\psid\rt\psi - \psid\vecna\psi\rqq
+ \frac{\hbar}{m}\vecna\times\lt\psid\sigbf\psi\rt\,,   \label{eq:CurrPauli}
$$
where $\psi$ is a Pauli 2-components spinor and $\sigbf$ is the usual Pauli
vector (2$\times$2) matrix. Also the above current appears as a sum of a
Newtonian part which, in the spinless (Newtonian) limit ($\hbar\sigbf\to
0$), is parallel to the classical momentum; and of a non-Newtonian spin
part due to the spin, which vanishes only for spinless bodies, {\em but
not in the NR limit}, i.e. for a small momentum $\imp$.
The Schr\"odinger
theory, a particular case of Pauli's, corresponds to constant spin,
$\sbf=\psid\sigbf\psi)/\psid\psi={\rm const.}$ Nevertheless the
Zitterbewegung term of current (\ref{eq:CurrPauli}) does not vanish
since the curl of quantity $\psid\sigbf\psi =
\psid\psi\sbf\equiv\rho\sbf$ is not zero (the probability density is in
general not constant). As the divergence of a curl vanishes identically,
also the Zitterbewegung term is conserved. Therefore it is
incorrect to assume, as usual, the Newtonian term as equal to the whole
Schr\"odinger current. This assumption leads to severe problems in the
interpretation of the nonvanishing mean kinetic energy for stationary
states described by {\em real} wavefunctions.} \cite{Salesi,Landau-1,Takabayasi}.

In a recent paper of ours \cite{NNM} it was proposed a {\em classical}
particle theory in which Zitterbewegung arises quite naturally.
For the above considerations we called that theory {\em Non-Newtonian
Mechanics} (NNM).
The classical motion of spinning particles was therein described without
recourse to particular models or special formalisms, and without employing
Clifford algebras, or classical spinors (appearing in all supersymmetric-like
classical models, as the Barut-Zanghi one \cite{Barut}), but simply by
generalizing the usual spinless theory.
Newtonian Mechanics is re-obtained as a particular case of that theory:
namely for spinless systems with no Zitterbewegung.

Let us remark that:\\
a) NNM does not fix either the dimension or the metric of the spacetime
(we could also have $D\neq 4$ representations of the quantized theory);\\
b) NNM (see below) is the most natural, straight  and economic
extension of the Newtonian Mechanics which allows a classical
theory of spin;\\
c) the Lagrangian and the equations of motion do not contain explicitly
the Lorentz factor: so that the ordinary (spinless) Lorentz transformations
represent special, not general, spacetime rotations [as a
consequence the velocity squared is not constrained to be equal to 1\cite{NNM}; cf.
also items i) and iii) below].

We start with the definition of the 4-velocity as the proper-time derivative
of the spacetime coordinate [the adopted metric is $(+;\;-,-,-)$]
\bb
v^\mu \equiv \dox^\mu \equiv \lt\frac{\dt}{\dtau};\;\frac{\rd\xbf}{\dtau}
\rt\,.                                        \label{eq:VDEF}
\ee
The proper time $\tau$ is defined as {\em the time measured in the
Center-of-Mass Frame} (CMF) where, by definition, the \mbox{3-momentum}
vanishes, $\imp=0$.
The reference system, where by definition the speed vanishes,
\mbox{$\vbf\equiv\rd\xbf/\dtau=0$}, is instead the {\em Rest Frame}
(RF) different from the CMF for spinning particles endowed with Zitterbewegung.
We assume also the on-shell constraint \ $p^2=m^2$, \ which implies, as shown in Ref.
\cite{NNM}, the ``Dirac constraint'' $p_\mu v^\mu=m$. \
In the absence of external fields, by only requiring spacetime isotropy and
homogeneity ($\doJ^\munu=\dop^\mu=0$), we get
\bb
\doS^\munu = -\doL^\munu = p^\mu v^\nu- p^\nu v^\mu\,.
\ee
After contracting the above equation with $p^\nu$ and exploiting the on-shell
constraint $p^2=m^2$, we obtain the Zitterbewegung equation:
\bb
v^\mu \ug \frac{p^\mu}{m} - \frac{\doS^\munu p_\nu}{m^2}\,. \label{eq:ZBWEq}
\ee
The global velocity contains a translational, time-constant, timelike
component $p^\mu/m$ related to the motion of the CM; and a rotational,
time-varying, spacelike \cite{NNM} component related to the
presence of the spin. We stress that very general conservation laws and
constraints are sufficient to derive the above equation of motion. The
consequent theory is the most general one and not a result of a particular
theoretical model.
Let us introduce the (non-conserved) 4-vector [3-vector $\kbf$ is the
Lorentz-boosts generator, $\kbf\equiv(S^{01},S^{02},S^{03})$]
\bb
\Wt^\mu
\equiv \frac{1}{m}S^\munu p_\nu =
\frac{1}{m}\lt\kbf\cdot\imp; \ p^0\kbf-\imp\times\sbf\rt
\ee
which in the CMF reduces to the boost 3-vector $\kbf$:
it is dual of the (conserved) spin Pauli-Lubanski 4-vector $W^\mu$
\bb
W^\mu
\equiv \frac{1}{m}\,\um\vare^{\mu\nu\rho\sigma}S_{\nu\rho}p_\sigma
= \frac{1}{m}\lt\sbf\cdot\imp; \ p^0\sbf-\imp\times\kbf\rt\,
\ee
which in the CMF reduces to the spin 3-vector $\sbf$.
Since the momentum is constant, we can re-write the Zitterbewegung
equation in a symmetric form
\bb
v^\mu \ug \frac{p^\mu}{m} - \frac{\dot{\Wt^\mu}}{m}\,. \label{eq:ZBwEq}
\ee
From Eq.\,(\ref{eq:ZBWEq}) it follows that in a generic frame
the trajectory is a helix around the constant direction of $\imp$.
We found in Ref. \cite{NNM} various consequences of the above Zitterbewegung equation, among
which:

i) luminal or superluminal ``global'' motions are allowed, without
violating Special Relativity, provided that the energy-momentum, and any
related signal or information, travel with a subluminal (average) speed;

ii) except for the case of polarized particles ($\dis\sbf\parallel\imp$),
the Zitterbewegung motion has a nonvanishing component along the
momentum;

iii) the ratio between the time durations measured in a generic inertial frame
and in the CMF is not constant and differs from the istantaneous Lorentz
factor which instead constitutes the mean value of $\rdt/\rd\tau$ on a {\em
Zitterbewegung} period. In general we can say that a non-linear relation
occurs between infinitesimal time durations measured in different inertial
frames.

We can build up, in a relativistically covariant way, also in the presence of
fields, a {\em Lagrangian} formulation of NNM through a straightforward
generalization of the Newtonian Lagrangian $\Lczero=\um mv^2$ to
Lagrangians containing time-derivatives of the velocity up to the $n$-th order:
\bb
\Lcn \equiv \um mv^2 + \um k_1\dov^2 + \um k_2\ddov^2 + \cdots - U
\equiv \sum_{i=0}^n\,\um k_i{v^{(\rm i)}}^2 - U\,,
\ee
where $U$ is a scalar potential due to external forces, the $k_i$ are
constant scalar coefficients endowed with alternate signs \cite{NNM}, $k_0=m$, and
\ $\vi\equiv\rd^{\rm i}v/\dtau^{\rm i}.$ \
The Euler-Lagrange equation of motion
$$
\frac{\pa\Lc}{\pa x} = \dot{\frac{\pa\Lc}{\pa\dox}} -
\ddot{\frac{\pa\Lc}{\pa\ddox}} + {\stackrel{\ldots}{\frac{\pa\Lc}{\pa\dddox}}} -
\cdots                                                  \label{eq:EL}
$$
gives a constant-coefficients $n$-th order differential equation, which
appears as a generalization of Newton's Law $F=ma$, in which non-Newtonian
Zitterbewegung terms appear:
\bb
-\,\frac{\pa U}{\pa x_\mu} \ug m\,a^\mu - k_1\,\ddoa^\mu +
k_2\,\ddddoa^\mu - \cdots\equiv
\sum_{i=0}^n\,(-1)^{{\rm i}}k_i\,\aii^\mu\,.     \label{eq:GNEq}
\ee
The canonical momentum
$\dis\frac{\pa\Lc}{\pa\dox_\mu} - \dot{\frac{\pa\Lc}{\pa\ddox_\mu}} +
\ddot{\frac{\pa\Lc}{\pa\dddox_\mu}} - \cdots$ \ conjugate to $x^\mu$
writes
\bb
p^\mu = m\,v^\mu - k_1\,\ddov^\mu + k_2\,\ddddov^\mu - \cdots
\equiv\sum_{i=0}^n\,(-1)^{{\rm i}}\,k_i\,{v^{(\rm 2i)}}^\mu\,, \label{eq:ZeroMom}
\ee
from which we get the Zitterbewegung equation for $\Lcn$:
\bb
v^\mu \ug \frac{p^\mu}{m} + \frac{k_1}{m}\,\ddov^\mu -
\frac{k_2}{m}\,\ddddov^\mu - \cdots = \frac{p^\mu}{m} -
\sum_{i=1}^n\,(-1)^{{\rm i}}\,\frac{k_i}{m}\,{v^{(\rm 2i)}}^\mu\,. \label{eq:LcnZBWeq}
\ee
Through the N\"other Theorem, by satisfying the symmetry under rotations,
the classical spin can be unequivocally defined employing only classical kinematical
quantities.
We get the spin tensor and the spin vector, respectively
(hereafter we refer to the first order theory; in Ref. \cite{NNM} it was
made also for $n>1$):
\bb
S^\munu\ug k_1\,\lt v^\mu a^\nu - v^\nu a^\mu\rt\,,
\ee
\bb
\sbf=k_1\,\lt\vbf\times\abf\rt\,.
\ee
The Hamiltonian representation of the theory is performed in NNM by
introducing, besides the zero order momentum $p^\mu$ given in
(\ref{eq:ZeroMom}), a first order momentum $\pi^\mu$ canonically conjugate to
$q^\mu\equiv v^\mu$:
\bb
\pi^\mu\equiv\frac{\pa\Lc}{\pa\doq_\mu}\equiv\frac{\pa\Lc}{\pa\dov_\mu}
=k_1a^\mu\,.
\ee
Thus the spin tensor can be expressed also in a canonical way:
\bb
S^\munu \ug q^\mu\pi^\nu - q^\nu\pi^\mu\,.
\ee
The Poisson brackets are here defined as follows
$$
\{f,g\} \equiv \left(\frac{\pa f}{\pa x_\mu}\frac{\pa g}{\pa p^\mu} -
\frac{\pa f}{\pa p_\mu}\frac{\pa g}{\pa x^\mu}\right) +
\left(\frac{\pa f}{\pa q_\mu}\frac{\pa g}{\pa\pi^\mu} -
\frac{\pa f}{\pa\pi_\mu}\frac{\pa g}{\pa q^\mu}\right)\,.
$$
The scalar Hamiltonian, which conserves because of the
$\tau$-reparametrization invariance, is
\bb
\Hc(\tau;\;x,p;\;q,\pi) \ug p_\mu\dox^\mu + \pi_\mu\doq^\mu - \Lc \ug
p_\mu q^\mu - \um mq^2 + \frac{\pi^2}{2k_1} + U\,.    \label{eq:Ham1Dirac}
\ee
The action $\Sc=\dis\int\Lc\dtau$ can be put in the characteristic form
\bb
\Sc \ug \int\,p_\mu\rd x^\mu + \pi_\mu\rd q^\mu - \Hc\dtau\,,
\ee
from which
\bb
p^\mu=\frac{\pa\Sc}{\pa x_\mu} \qquad\qquad
\pi^\mu=\frac{\pa\Sc}{\pa q_\mu} \qquad\qquad
\Hc=-\,\frac{\pa\Sc}{\dtau}\,.
\ee
Besides the standard couple of Hamilton equations, we have a non-Newtonian
couple of Hamilton equations, applying to the second order pair of canonical
variables $(q^\mu,\;\pi^\mu)$
\bb
\left\{\begin{array}{l}
{\dis\frac{\pa\Hc}{\pa p_\mu} \ug \dox^\mu}\\
\ \\
{\dis\frac{\pa\Hc}{\pa x_\mu} \ug -\dop^\mu}
\end{array}\right.
\qquad\qquad\qquad
\hfill\left\{\begin{array}{l}
{\dis\frac{\pa\Hc}{\pa\pi_\mu} \ug \doq^\mu}\\
\ \\
{\dis\frac{\pa\Hc}{\pa q_\mu} \ug -\dopi^\mu} \ \ .
\end{array}\right.
\ee
(for $n>1$ we have to introduce other higher order momenta up to
the $n$-th order entering $n+1$ systems of pairs of Hamilton
equations).
As shown in Ref. \cite{NNM}, the above Hamilton equations are fully equivalent to
the Euler-Lagrange equation, that is to generalized Newton's Law
(\ref{eq:GNEq}).

\section{Three-dimensional and non-relativistic theory}

\noindent Let us fix the only (apart the mass $m$) NNM
free parameter $k_1$ to $-\hbar^2/4mc^4$,
so that (considering for simplicity free particles, $U=0$):
\bb
\pi^\mu = - \frac{\hbar^2}{4mc^4}a^\mu
\ee
\bb
\Lc \ug \um\,m\,v^2 - \frac{\hbar^2}{8mc^4}a^2           \label{eq:LagrDirac}
\ee
\bb
\Hc \ug pq-\um\,mq^2-\frac{2mc^4}{\hbar^2}\pi^2                     \label{eq:HamDirac}
\ee
\bb
S^\munu = \frac{\hbar^2}{4mc^4}(a^\mu v^\nu - a^\nu v^\mu)\,.  \label{eq:SpinTensor}
\ee
Extending NNM to macroscopic bodies we find, as expected,
a Newtonian behavior with vanishing spin and Zitterbewegung because of
the extreme smallness of $k_1=-\hbar^2/4mc^4$ when $m\to\infty$.

The Zitterbewegung equation of motion (\ref{eq:LcnZBWeq}) now reduces to
\bb
v^\mu \ug \frac{p^\mu}{m} - \frac{\hbar^2}{4m^2c^4}\ddov^\mu\,,    \label{eq:ZBWEqMNN1}
\ee
whose general solution oscillates with the so-called ``Compton frequency''
$\omega_{\rm c}=2mc^2/\hbar$ ($E^\mu$, $H^\mu$ are constant spacelike 4-vectors fixing the
``internal'' initial conditions whilst $p^\mu$ fixes the ``external'' one)
\bb
v^\mu = \frac{p^\mu}{m} + E^\mu\cos(\omega_{\rm c}\tau) +
H^\mu\sin(\omega_{\rm c}\tau)\,.
\ee
The Hamilton generalized equations in NNM, globally equivalent to
Eq.\,(\ref{eq:ZBWEqMNN1}), become
\bb
\left\{\begin{array}{l}
{\dis\frac{\pa\Hc}{\pa p_\mu} = q^\mu = \dox^\mu}\\
\ \\
{\dis\frac{\pa\Hc}{\pa x_\mu} = 0 = -\dop^\mu}
\end{array}\right.
\qquad\qquad\qquad
\hfill\left\{\begin{array}{l}
{\dis\frac{\pa\Hc}{\pa\pi_\mu} = -\frac{4mc^4}{\hbar^2}\pi^\mu = \doq^\mu}\\
\ \\
{\dis\frac{\pa\Hc}{\pa q_\mu} = p^\mu - mq^\mu = -\dopi^\mu} \ \ .
\end{array}\right.
\ee
As a consequence, also the second-order canonical variables $q$ and
$\pi$ are harmonic oscillator coordinates with the Compton frequency
\bb
\ddoq^\mu +\omega_{\rm c}^2q^\mu = 4mp^\mu
\qquad\qquad \ddopi^\mu +\omega_{\rm c}^2\pi^\mu = 0\,.
\ee
In Appendix it is shown that NNM is strictly
related to the {\em Proper Time} Dirac theory\cite{Dirac,NNM,Staunton1},
a representation of SO(3,2) Lie algebra.
Actually, some quantum Heisenberg equations
$\widehat{\dot{G}}=i\,[\Hco,\;\Go]$ have classical counterparts in
the Poisson brackets equations \ $\dot{G}=\{\Hc,\;G\}$ \ if we take the
operator $\Hco$ equal to the scalar Hamiltonian $\po_\mu\ga^\mu\!-\!m$
in Proper-Time Dirac theory\footnote{The very Dirac equation
$\po_\mu\ga^\mu\!-\!m=0$ arises by taking $\Hco$, according to Dirac's definition, ``weakly
zero''.} and $\Hc$ equal to NNM Hamiltonian, Eq.\,(\ref{eq:HamDirac}).

\

\

\noindent Just as it occurs in the NR limit of the Dirac theory (i.e. in quantum
Pauli's and Schr\"odinger's theories), we have a classical {\em
Zitterbewegung} motion also in the NR limit of NNM. Actually
we have a motion even in the CMF. In fact, for $\imp\to 0$ we have
$\doS^{ik}\to 0$ (since the spin 3-vector conserves in NR Mechanics) but
$\doS^{i0}{\to\!\!\!\!\!\!/} \ \ 0$ ($S^{i0}$ is not required to conserve), so
that from Eq.\,(\ref{eq:ZBWEq}) we have $v^i\to-\doS\cmf^{i0}/m\neq 0$.
Therefore it is physically meaningful to study the NR NNM.

To this end it is sufficient to recall that in the NR limit
the ordinary time coordinate $t$ reduces to the proper time $\tau$:
so that all the time derivatives in Zitterbewegung equation
(\ref{eq:ZBWEqMNN1}) can now be meant as taken with respect to
$t$.

From the spatial part of  (\ref{eq:ZBWEqMNN1}) we
get the 3-momentum of a free particle
\bb
\imp \ug m\vbf+\frac{\hbar^2}{4mc^4}\doabf\,. \label{eq:3-momentum}
\ee
In the presence of an external force $\Fbf$ and of a potential $U(\xbf)$, we can
write
\bb
\frac{\rd\imp}{\dt} = -\vecna U = \Fbf\,,
\ee
an then, by time-differentiating Eq.\,(\ref{eq:3-momentum}),
\bb
\fbox{${\dis\Fbf \ug m\abf + \frac{\hbar^2}{4mc^4}\,\ddoabf}$}
\label{eq:GNL}
\ee

\noindent where it appears a non-Newtonian term
which becomes important only when $\ddoabf$ is very large.

\

\

\noindent Let us derive the interesting non-Newtonian extension of the
Work-Kinetic Energy Theorem holding in classical Newtonian mechanics
\ $\displaystyle T = \int\Fbf\cdot\rd\xbf =  \frac{1}{2}\,m\vbf^2\,.$ \
We now have
\bb
T = \int\Fbf\cdot\rd\xbf =
\int\lt m\abf+\frac{\hbar^2}{4mc^4}\ddoabf\rt\cdot\rd\xbf =
\int\lt m\abf+\frac{\hbar^2}{4mc^4}\ddoabf\rt\cdot\vbf\dt\,.
\ee
Taking in account that identically $\dis\abf\cdot\vbf=\frac{\rd}{\dt}
\lt\frac{\vbf^2}{2}\rt$ and
$\dis\ddoabf\cdot\vbf=-\,\frac{\rd}{\dt}\lt\frac{\abf^2}{2} -
\doabf\cdot\vbf\rt$
we can write a non-Newtonian Work-Kinetic Energy Theorem where, besides
the usual Newtonian term, a term appears which depends on the first two
derivatives of the velocity
\bb
\fbox{${\dis T \ug \frac{1}{2}\,m\vbf^2 -
\frac{\hbar^2}{4mc^4}\,\lt\frac{\abf^2}{2} - \doabf\cdot\vbf\rt}$}\,.
\ee
It is easy to show that the total mechanical energy $T+U$ is a conserved quantity
because turns out to be equal, up to a constant, to the 3$D$ NR limit of
Hamiltonian (\ref{eq:Ham1Dirac}), that is
\ $\imp\cdot\qbf-\um m\qbf^2-2m\pibf^2+U$.

\

\noindent There is a noticeable consequence of this theorem: the non-existence
of the classical ``potential barrier'',
the classical analogue of the quantum tunnel effect.
In fact from the conservation of the energy
\bb
E  = T + U = \frac{1}{2}\,m\vbf^2 - \frac{\hbar^2}{4mc^4}\,\lt\frac{\abf^2}{2} -
\doabf\cdot\vbf\rt + U
\ee
we see that, even in the space regions where $E<U$, quantity $\vbf^2$ can
be positive, for the counterbalancing
presence in $T$ of the non-Newtonian term.

Let us also remark the analogy between the conservation equation
\bb
E  = \frac{1}{2}\,m\vbf^2 - \frac{\hbar^2}{4mc^4}\,\lt\frac{\abf^2}{2} -
\doabf\cdot\vbf\rt + U
\ee
and the ``quantum Hamilton-Jacobi'' equation for the Madelung fluid of the
Schr\"odinger wave-Mechanics \ ($\dis\psi\equiv\sqrt{\rho}\erm^{i\varphi/\hbar}$) \
\bb
- \pa_{t}\varphi = \frac{1}{2m}\lt\vecna\varphi\rt^2
+ \frac{\hbar^2}{4m}\lqq\um\lt\frac{\vecna\rho}{\rho}\rt^2
- \frac{\triangle\rho}{\rho}\rqq + U\,.
\ee
Actually, we are induced to relate the quantum phase and the classical action
\bb
-\pa_t\varphi \ \Leftrightarrow \ E=-\pa_t\Sc
\qquad\qquad
-\frac{\vecna\varphi}{m} \ \Leftrightarrow \ \vbf=\frac{\vecna\Sc}{m}
\ee
as well as the  so-called ``quantum potential'' and the non-Newtonian kinetic
term
\bb
\frac{\hbar^2}{4m}\lqq\frac{1}{2}\lt\frac{{\vecna}\rho}{\rho}\rt^2
- \frac{\triangle\rho}{\rho}\rqq
\ \Leftrightarrow \ -\frac{\hbar^2}{4mc^4}\lt\um\abf^2-\doabf\cdot\vbf\rt\,.
\ee

\

\

{\bf Acknowledgements}

\noindent The Author is glad to thank G.\,Andronico, G.G.N.\,Angilella,
F.\,Bottacin, L.\,Brandolini, M.\,Consoli, M.\,Villa,
D.\,Zappal\`a and, in particular, S.\,Esposito and E.\,Recami.

\

\appendix

\ \ {\large\bf Appendix --- {\em Correspondences between NNM and Proper-Time Dirac
Theory}}\\

\noindent For $k_1=-1/4m$ (hereafter we adopt, as usual, $\hbar=c=1$) NNM
is strictly related to the Proper-Time Dirac theory.
As a matter of fact we can find exact correspondences between quantum
and classical time-evolution equations. That is, between the Heisenberg
equations $\widehat{\dot{G}}=i\,\lqq\Hco,\;\Go\rqq$
and the Poisson brackets equations \ $\dot{G}=\{\Hc,\;G\}$
(taking $\Hco=\po_\mu\ga^\mu\!-\!m$ and
$\dis\Hc=pq-\um\,mq^2-2m\pi^2$).
For example we have
\bb
\widehat{\dop}\,^\mu = i\,\lqq\Hco,\;\po^\mu\rqq = 0
\ee
\bb
\widehat{\doS}\,^\munu = i\,\lqq\Hco,\;\So^\munu\rqq = \po^\mu\ga^\nu
- \po^\nu\ga^\mu\,;
\ee
which, taking in account that $\ga^\mu=i\lqq\Hco,x^\mu\rqq=\vo^\mu$,
correspond to the classical equations of conservation of the linear and
angular momenta
\bb
\dop^\mu = \{\Hc,\;p^\mu\} = 0                \label{eq:pdot}
\ee
\bb
\doS^\munu = \{\Hc,\;S^\munu\} = p^\mu v^\nu - p^\nu v^\mu\,. \label{eq:Sdot}
\ee
Furthermore we have for the 4-acceleration operator
\bb
\aoo^\mu = i\,\lqq\Hco,\;\vo^\mu\rqq =  4\,\So^{\mu\nu}\po_\nu\,;
\ee
correspondingly, in NNM, by contracting both sides of (\ref{eq:SpinTensor})
with $p_\nu$ and exploiting the on-shell constraint $p^2=m^2$, we have
\bb
S^\munu p_\nu = \frac{a^\mu}{4}\,.
\ee
Notice that, differently from equations (\ref{eq:pdot}), (\ref{eq:Sdot}),
the last equation holds only in \nnmuno and not for larger $n$ anymore.

Besides the above correspondences, let us now recover in Proper Time
Dirac theory just the NNM Zitterbewegung equation,
Eq.\,(\ref{eq:ZBWEqMNN1}). Let us apply the Heisenberg equation to the
4-acceleration operator $\aoo^\mu$
\bb
\doao^\mu = i\,\lqq\Hco,\;\aoo^\mu\rqq =
i\,\lqq\po_\lambda\ga^\lambda\!-\!m,\;4\,\So^{\mu\nu}\po_\nu\rqq =
-4\po^2\ga^\mu + 4\po^\mu\,\po_\lambda\ga^\lambda\,,
\ee
and consider only the physical on-shell states which satisfy both
Dirac and Klein--Gordon equation, \ $\po_\lambda\ga^\lambda\psi=m\psi$ and
$\po^2\psi=m^2\psi$. \ When applying to the vectors of that Hilbert
subspace we can write
\bb
\doao^\mu \ug -4m^2\vo^\mu + 4m\po^\mu\,,
\ee
which can be put just in the form of
an operator version of the Zitterbewegung equation, Eq.(\ref{eq:ZBWEqMNN1}):
\bb
\vo^\mu \ug \frac{\po^\mu}{m} - \frac{\doao^\mu}{4m^2}\,.
\ee

\

\end{document}